\documentclass[a4paper,12pt]{article}

\newcommand{\sect}[1]{\setcounter{equation}{0}\section{#1}}

\newcommand{\bea}{\begin{eqnarray}}
\newcommand{\eea}{\end{eqnarray}}
\newcommand{\be}{\begin{equation}}
\newcommand{\ee}{\end{equation}}
\newcommand{\vs}[1]{\vspace{#1 mm}}

\newcommand{\dsl}{\pa \kern-0.5em /}

\newcommand{\pa}{\partial}

\begin{document}
\topmargin 0pt
\oddsidemargin 0mm

\begin{flushright}
hep-th/0205198\\
\end{flushright}

\vs{2}
\begin{center}
{\Large \bf  
On supergravity solutions of space-like D$p$-branes}
\vs{10}

{\large Shibaji Roy}
\vspace{5mm}

{\em 
 Saha Institute of Nuclear Physics,
 1/AF Bidhannagar, Calcutta-700 064, India\\
E-Mail: roy@theory.saha.ernet.in\\}
\end{center}

\vs{5}
\centerline{{\bf{Abstract}}}
\vs{5}
\begin{small}
Recently the time dependent solutions of type II supergravities in $d = 10$,
with the metric having the symmetry $ISO(p+1) \times SO(8-p, 1)$ have been
given by two groups (Chen-Gal'tsov-Gutperle (CGG), [hep-th/0204071] and
Kruczenski-Myers-Peet (KMP), [hep-th/0204144]). The supergravity solutions
correspond to space-like D$p$-branes in type II string theory. While the
CGG solution is a four parameter solution, the KMP solution is a three
parameter solution and so in general they are different. This difference can
be attributed to the fact that unlike the CGG solution, KMP uses a specific 
boundary condition for the metric and the dilaton field. It is shown that
when we impose the boundary conditions used in the KMP solution to the CGG
solution then both become three parameter solutions and they map to each 
other under a coordinate transformation along with a Hodge duality of the 
field strength. We also give the relations between the parameters 
characterizing the two solutions.
\end{small}

\sect{Introduction}

Recently there has been a lot of interest in constructing and understanding
the time dependent solutions in string/M theory. The major motivations for
studying these solutions are: (a) they might provide the stringy resolution
of space-like cosmological singularity behind the black hole horizon,
(b) they might provide a concrete realization of dS/CFT correspondence 
\cite{as,bdm} in
string theory. The issue of singularity has been addressed in the context
of time dependent orbifold model in string theory in \cite{kosst,bskn,ne,cc,
cck}. 
Also many physical
issues like observables, perturbation theory, particle creation have been 
addressed in a simple time dependent string background in \cite{afhs,lms}.

Space-like $p$-branes (S$p$-branes) are topological defects localized in 
$(p+1)$ dimensional space-like surfaces and are known to arise in string/M
theory (also in some field theories) as time dependent solutions \cite{gs}. 
Space-like
D$p$-branes (SD$p$-branes) arise when the time-like as well as $(8-p)$
space-like coordinates of the open string satisfy Dirichlet boundary 
condition \cite{hu,huone} and carry the same kind of RR charge as the ordinary 
(time-like) D$p$-branes.
They can also be understood \cite{asen} to arise from unstable 
D$(p+1)$-brane or
ordinary D$p$-brane-anti-D$p$-brane pair as the time-like tachyonic kink
solution. The supergravity description of the SD$p$-branes are particularly
interesting to understand the time-like holography principle of the dS/CFT
correspondence \cite{asone}.

The supergravity description of SD$p$-branes has been given by 
Chen-Gal'tsov-Gutperle (CGG) \cite{cgg} and also by Kruczenski-Myers-Peet 
(KMP) \cite{kmp}. These
two solutions look quite different and in fact the methods used to obtain 
these two solutions are also completely different. In the first case CGG
started with a coupled dilaton, gravity and a $q$-form field strength system
in $d$ space-time dimensions which is the bosonic sector of low energy
effective action of various string theories or M theory compactified to
$d$ dimensions. The non-linear differential equations resulting from the
effective action are then solved with an ansatz for the metric to have the
symmetry structure $ISO(p+1) \times SO(d-p-2, 1)$. The resulting time 
dependent solution
for $d = 10$ and the dilaton coupling $a = (p-3)/2$ represents the localized
(we take $k=q$ in the CGG solution and for $d=10$, $q=8-p$) SD$p$-branes of 
type II string theory \cite{cgg}. These
are magnetically charged SD$p$-branes in the Einstein-frame metric and are
dependent on four parameters. On the other hand, KMP started with the eleven
dimensional solution of the equations of motion of pure Einstein gravity
with appropriate symmetries \cite{cmp}. Then they performed a 
rotation mixing the 
eleventh dimension and one of the space-like dimensions. The dimensional
reduction of the eleventh dimension produces the SD0-brane solution of 
type IIA string theory smeared in some number of transverse directions. The
usual solution generating technique of T-duality \cite{bho,bmm} in the 
transverse directions
then  gives the required electric SD$p$-brane solutions. Demanding isotropy
in $(p+1)$ directions gives a three parameter solution of SD$p$-branes in the
string frame \cite{kmp}.

Since the CGG solution is a four parameter solution, whereas the KMP solution
is a three parameter solution, they are in general different. The purpose
of this paper is to show under what condition they will be the same as both
of them represent the SD$p$-brane supergravity solutions. In fact we will
point out that unlike the CGG solution, the KMP solution uses specific
boundary conditions for the metric as well as for the dilaton. We show that
when the same boundary conditions used by the KMP solution is imposed upon
the CGG solution, then the two solutions indeed map to each other\footnote{
A similar mapping for the static solution has been pointed out in \cite{cgg}
by comparing their solution with that of \cite{bmo}.} under a
coordinate transformation
along with a Hodge duality of the field strength (since in the CGG case 
SD$p$-branes are magnetically charged whereas in the KMP case they are
electrically charged). The coordinate transformation has the form:
\bea
\frac{t}{\omega} &=& \left[\tanh \frac{(7-p)\hat{t}}{2}\right]^{-
\frac{1}{7-p}}\nonumber\\
{\rm or},\quad  \hat{t} &=& - \frac{1}{7-p} \ln \left[\frac{1 - 
(\frac{\omega}{t})^{7-p}}{1 + (\frac{\omega}{t})^{7-p}}\right]
\eea
where $t$ is the time-like coordinate in the KMP solution and $\omega$ is a 
parameter and $\hat{t}$ is the time-like coordinate in CGG. 
Note that as $t \to \infty$,
$\hat{t} \to 0$ and as $t \to \omega$, $\hat{t} \to \infty$. We will also
give the relations between the parameters characterizing the two solutions.
If we demand that the string-frame metric becomes flat (in Rindler
coordinates) and $e^{2\phi}$ approaches unity for $t \to \infty$
(or $\hat{t} \to 0$), to the CGG solution (as has been used by the 
KMP solution), then both the solutions become three parameter solution
and they map to each other under the above coordinate transformation along 
with a Hodge duality of the field strength. In fact, when we use these boundary
conditions then one of the parameters in the CGG solution gets related to the
other three paramaters in a specific way. This is the reason that in this 
case both the solutions become three parameter solutions which is necessary 
for the complete mapping of these two solutions. 
Thus we show that under these circumstances the CGG solution and the
KMP solution become identical to each other.
 
The organization of this paper is as follows. In section 2, we compare the
space-like M5-brane solutions obtained by Gutperle-Strominger (GS) and KMP
and fix our notations.
In section 3, we discuss the mapping of SD$p$-brane solutions of CGG and KMP.
We summarize our conclusion in section 4.

\sect{SM5-brane solutions}

In this section we will discuss the equivalence of SM5-brane solutions 
obtained by
GS \cite{gs} and KMP \cite{kmp}. We will point out that in both cases 
the SM5-brane solutions are 
two-parameter solutions and they map to each other under a coordinate 
transformation as well as a Hodge duality of the field strength. In \cite{gs}
the SM5-brane solution is obtained by solving the equation of motion resulting
from the bosonic action of $d = 11$ supergravity. By imposing the appropriate
symmetry the supergravity solution of SM5-brane is found to have the form,
\bea
ds^2 &=& - e^{2A} d\hat{t}^2 + e^{2C} dH_{d-p-2}^2 + e^{2B} dx_{(p+1)}^2
\nonumber\\
\ast F_{p+2} &=& e^{2B(p+1)} b\,\, d\hat{t} \wedge dx_1\wedge \ldots \wedge 
dx_{p+1}
\eea
where $A$, $B$, $C$ are functions of $\hat{t}$ only and satisfy the gauge 
condition
\be
-A + (p+1) B + (d-p-2) C = 0
\ee
We have written the solution in (2.1) such that it can be generalized for 
SD$p$-branes to be discussed in the next section. In this particular case of 
SM5-brane solution $p=5$ and $d=11$. $dH_{d-p-2}^2$ is the 
line element for the
$(d-p-2)$-dimensional hyperbolic space with negative curvature and 
$dx_{(p+1)}^2$ is the same for the flat $(p+1)$-dimensional Euclidean space.

Since the functions $A$, $B$, $C$ satisfy the gauge condition (2.2), they
can be expressed in terms of two functions as follows,
\be
A = (d-p-2) g(\hat{t}) - \frac{(p+1)}{(d-p-3)} f(\hat{t}), \qquad 
B = f(\hat{t}), \qquad C = g(\hat{t}) - \frac{(p+1)}{(d-p-3)} f(\hat{t})
\ee
We will use these forms for the SD$p$-brane solution of type II string 
theory in
the next section for $d=10$.
Solving the equations of motion the functions $g$ and $f$ are found to have
the forms\footnote{These are actually solutions of $d$-dimensional gravity
coupled to $(d-p-2)$-form field strength system.}
\bea
f(\hat{t}) &=& \frac{2}{\chi} \ln \frac{\alpha}{\cosh\frac{\chi\alpha}{2}
(\hat{t} - t_0)} + \frac{1}{\chi} \ln \frac{(d-2) \chi}{(d-p-3) b^2}
\nonumber\\
g(\hat{t}) &=& \frac{1}{d-p-3} \ln \frac{\beta}{\sinh (d-p-3) \beta (\hat{t}
-t_1)}
\eea
where $\chi = 12$ in this case and $\alpha$, $\beta$ are integration 
constants satisfying
\be
\frac{(d-2) \chi \alpha^2}{2(d-p-3)} - (d-p-2)(d-p-3) \beta^2 = 0
\ee
Also, $t_0$ and $t_1$ are two other integration constants. Note that $t_1$
can be absorbed by shifting $\hat{t}$ coordinate. For SM5 solution (2.5) 
reduces to
\be
3\alpha^2 - 2\beta^2 = 0
\ee
 and the functions $f(\hat{t})$ and $g(\hat{t})$ simplify to \cite{gs}
\bea
f(\hat{t}) &=& \frac{1}{6} \ln \frac{\beta}{\cosh \sqrt{24} \beta (\hat{t}-
t_0)} - \frac{1}{12} \ln (\frac{b^2}{24})\\
g(\hat{t}) &=& \frac{1}{3} \ln \frac{\beta}{\sinh 3\beta\hat{t}}
\eea
where we have eliminated the integration constant $\alpha$ using (2.6).
Furthermore, the constant $\beta$ can also be eliminated by scaling
$\hat{t} \to \hat{t}/\beta$ and $x_i \to x_i/\beta^{1/6}$ for $ i=1, \ldots,
6$. We remark here that $\beta$ can not be eliminated by similar rescaling
of coordinates only for SD$p$ solutions to be discussed later. Thus 
eliminating $\beta$ for this case the metric and the 7-form dual field
strength take the following forms,
\bea
ds^2 &=& \left(\frac{b^2}{24}\right)^{1/3} \frac{\left(\cosh 
\sqrt{24}(\hat{t} - t_0)
\right)^{2/3}}{(\sinh 3\hat{t})^{8/3}}\left( - d\hat{t}^2 + \sinh^2 3\hat{t}
dH_4^2\right)\nonumber\\
& &\qquad + \left(\frac{b^2}{24}\right)^{-1/6}\frac{1}{(\cosh\sqrt{24}
(\hat{t}-t_0))^{1/3}}
dx_{(6)}^2\nonumber\\
\ast F_7 &=& e^{12 f} b \,\, d\hat{t}\wedge dx_1\wedge \ldots \wedge dx_6
\eea
The above equation represents the SM5-brane solution characterized by 
two parameters
$b$ and $t_0$. We now make a coordinate transformation
\be
\hat{t} = - \frac{1}{3} \ln\frac{f_-}{f_+}
\ee
where
\be
f_{\pm} = 1 \pm \left(\frac{\omega}{t}\right)^3
\ee
Then we find,
\be
\frac{1}{(\sinh 3\hat{t})^{8/3}}\left( - d\hat{t}^2 + \sinh^2 3\hat{t} dH_4^2
\right) = \frac{(f_+ f_-)^{2/3}}{2^{2/3} \omega^2}(-dt^2 + t^2 dH_4^2)
\ee
and we rewrite
\bea
\left(\frac{b^2}{24}\right)^{1/2} \cosh \sqrt{24}(\hat{t} - t_0)
&=& 2\omega^3 \Big[\cos^2\theta \left(\frac{f_-}{f_+}\right)^{-\sqrt{8/3}}
+ \sin^2\theta \left(\frac{f_-}{f_+}\right)^{\sqrt{8/3}}\Big]\nonumber\\
&=& 2\omega^3 F
\eea
where we have defined
\bea
2\omega^3 \cos^2\theta &=& \frac{1}{2} e^{-\sqrt{24} t_0}\left(\frac{b^2}{24}
\right)^{1/2}\nonumber\\
2\omega^3 \sin^2\theta &=& \frac{1}{2} e^{\sqrt{24} t_0}\left(\frac{b^2}{24}
\right)^{1/2}
\eea
Note here that $\theta$ is the mixing angle of the eleventh dimension
and one of the space-like dimensions used in \cite{kmp} to construct the 
SM5-brane solution.
Now using (2.12) -- (2.14), the metric and the field strength in (2.9)
can be rewritten as,
\bea
ds^2 &=& F^{2/3} \left(f_+ f_-\right)^{2/3} \left[ - dt^2 + t^2 dH_4^2\right]
+ F^{-1/3} dx_{(6)}^2\nonumber\\
F_4 &=& -b\, \epsilon(H_4) = -6 \sin\theta \cos\theta (2 \sqrt{\frac{8}{3}})
\omega^3 \epsilon(H_4)
\eea
This is precisely the same form of the metric and the field strength obtained
in \cite{kmp} for the SM5-brane solution. Note that the coordinates 
$x_i$ for $i=1,2,
\ldots,6$ are rescaled while we write (2.15) by $x_i \to (2\omega^3)^{1/6}
x_i$. Also we have taken the Hodge dual of the field strength in (2.9) to
write the first expression of $F_4$ in (2.15). The second expression is 
written using (2.14). $\epsilon(H_4)$ represents the volume form of 
4-dimensional hyperbolic space. So both the GS solution and the KMP solution 
are two
parameter solutions. The relations between the GS parameters ($t_0$, $b$)
and the KMP parameters ($\omega$, $\theta$) follow from (2.14) as,
\bea
\omega^3 &=& \frac{1}{2} \left(\frac{b^2}{24}\right)^{1/2} \cosh \sqrt{24} t_0
\qquad \Rightarrow \qquad b = 4\sqrt{24} \omega^3 \sin\theta 
\cos\theta\nonumber\\
\tan \theta &=& e^{\sqrt{24} t_0} \qquad \Rightarrow \qquad t_0 = \frac{1}
{\sqrt{24}}\ln(\tan\theta)
\eea
Thus we have shown the exact mapping of the two apparently different looking
solutions of SM5-brane obtained in \cite{gs} and \cite{kmp} by the 
coordinate transformation
(2.10).

\sect{SD$p$-brane solutions}

In this section we show the equivalence of the SD$p$-brane solutions 
obtained in \cite{cgg} and \cite{kmp} along the same line as in the 
previous section.
We point out that the CGG solution is a four parameter solution and the
KMP solution 
is a three parameter solution. The difference is because the KMP solution
uses a specific boundary condition that the string-metric in their solution
becomes flat in Rindler coordinates and $e^{2\phi}$ approaches unity
as $t \to \infty$ (or, $\hat{t} \to 0$), whereas for the CGG solution 
they remain arbitrary. This
arbitrariness shows up as an additional parameter in the CGG solution.
However, as we impose these additional restrictions in the CGG solution, 
we find that one of the parameters in the CGG solution is removed and then
these two solutions map to each other.
Let us start with the CGG solution which has
the form as given in (2.1) along with the gauge condition (2.2) where now
$d=10$ and there is a dilaton $\phi(\hat{t})$. Solving the equations
of motion the functions $f(\hat{t})$, $g(\hat{t})$ and $\phi(\hat{t})$ were
obtained in \cite{cgg} to have the forms,
\bea
f(\hat{t}) &=& \frac{2}{\chi} \ln \frac{\alpha}{\cosh\frac{\chi\alpha}{2}
(\hat{t} - t_0)} + \frac{1}{\chi} \ln \frac{8 \chi}{(7-p) b^2} -
\frac{(p-3)c_1}{2\chi}\hat{t} - \frac{(p-3)c_2}{2\chi}
\nonumber\\
g(\hat{t}) &=& \frac{1}{7-p} \ln \frac{\beta}{\sinh(7-p) \beta (\hat{t}
-t_1)}\nonumber\\
\phi(\hat{t}) &=& \frac{4(p-3)}{(7-p)} f(\hat{t}) + c_1 \hat{t} 
+ c_2\nonumber\\
{\rm and}\quad F_{8-p} &=& b \epsilon(H_{8-p})
\eea
In the above $\chi = 32/(7-p)$ and $\alpha$, $\beta$, $t_0$, $t_1$, 
$c_1$ and $c_2$ are the integration constants. $b$ is related to the 
magnetic charge of the solution. We remark that $c_1$ and $c_2$ are two more
integration constants (than in the previous case) which appeared while
solving the dilaton equation of motion. 
The constants $\alpha$, $\beta$ and $c_1$ 
are related by,
\be
\frac{(p+1)}{\chi}c_1^2 + \frac{4\chi}{(7-p)}\alpha^2 - (8-p)(7-p)\beta^2
= 0
\ee
Note that we can absorb $t_1$ by shifting $\hat{t}$ and therefore the solution
depends on six parameters $\beta$, $t_0$, $c_1$, $c_2$ and $b$ with a relation 
between $\alpha$, $\beta$ and $c_1$ given in (3.2). 
Now in order 
to show the mapping we write the full CGG solution of SD$p$-branes using (3.1)
and (2.1) -- (2.3) as,
\bea
ds^2 &=&\left(\frac{\beta}{\sinh(7-p)\beta\hat{t}}\right)^
{\frac{2(8-p)}{(7-p)}}\left(\cosh\frac{\chi\alpha}{2}(\hat{t}-t_0)\right)^{1/2}
\left(\frac{(7-p)b}{16\alpha}\right)^{1/2} e^{\frac{p+1}{8}(c_1\hat{t}+c_2)}
\nonumber\\
& & \qquad\qquad\qquad\qquad\qquad\qquad \times \left[-d\hat{t}^2 + 
\frac{\sinh^2(7-p)\beta\hat{t}}{\beta^2} dH_{8-p}^2\right]
\nonumber\\
& & + \left(\cosh\frac{\chi\alpha}{2}(\hat{t}-t_0)\right)^{-1/2}
\left(\frac{(7-p)b}{16\alpha}\right)^{-1/2} e^{\frac{7-p}{8}(c_1\hat{t}+c_2)}
dx_{(p+1)}^2\nonumber\\
e^{2\phi} &=& \left(\cosh\frac{\chi\alpha}{2}(\hat{t}-t_0)\right)^
{\frac{3-p}{2}}\left(\frac{(7-p)b}{16\alpha}\right)^{\frac{3-p}{2}}
e^{\frac{(p+1)(7-p)}{8}(c_1\hat{t}+c_2)}\nonumber\\
F_{8-p} &=& b \epsilon(H_{8-p})
\eea
Here we have written the metric in the string frame by 
multiplying the expression of (2.1) with $e^{\phi/2}$ since the KMP metric
is given in the string frame. Also in order to compare with the KMP solution
we have to dualize the field strength. Now we would like to point out
that in the above solution $\beta$ can not be eliminated by just rescaling
the coordinate $\hat{t}$ and $x_i$, for $i=1,\ldots,p+1$, as has been done
for the SM5-brane in the previous section. However, we find that there
is a unique way one can eliminate $\beta$ and this is done by renaming
the parameters as follows,
\bea
\frac{\alpha}{\beta} &\rightarrow& \alpha\nonumber\\
\beta t_0 &\rightarrow& t_0\nonumber\\
\frac{c_1}{\beta} &\rightarrow& c_1\nonumber\\
\beta^{\frac{p-3}{2}} e^{\frac{(p+1)(7-p)}{8}c_2} 
&\rightarrow& e^{\frac{(p+1)(7-p)}
{8}c_2} \nonumber\\
b &\rightarrow& b
\eea
alongwith the coordinate rescaling $\hat{t} \rightarrow \hat{t}/\beta$
and $x_i \to x_i/\beta^{1/(p+1)}$, for $i=1,\ldots,p+1$. 
Note that the renaming of the parameter
$c_2$ is not necessary for $p=3$. In fact, in this case
the renaming of other parameters and
the coordinate rescalings is enough to eliminate $\beta$ completely from the
solution (3.3). As $p=3$ case is different from the other cases we will
discuss the mapping for this case at the end of this section. Also, it
should be emphasized that only when $\beta$ is eliminated $e^{2\phi}$ 
approaching unity and the string metric becoming flat as $\hat{t} \to 0$ 
can be achieved.
Now the solution depends on five parameters $\alpha$, $t_0$,
$c_1$, $c_2$, $b$ with a relation between $\alpha$ and $c_1$ of the form
(see eq.(3.2))
\be
\frac{(p+1)}{\chi}c_1^2 + \frac{4\chi}{(7-p)}\alpha^2 = (8-p)(7-p)
\ee
Therefore if we eliminate one of $c_1$ and $\alpha$, then the solution would
depend on four parameters. Eliminating $\beta$ the solution (3.3) reduces
to,
\bea
ds^2 &=&\left(\frac{1}{\sinh(7-p)\hat{t}}\right)^
{\frac{2(8-p)}{(7-p)}}\left(\cosh\frac{\chi\alpha}{2}(\hat{t}-t_0)\right)^{1/2}
\left(\frac{(7-p)b}{16\alpha}\right)^{1/2} e^{\frac{p+1}{8}(c_1\hat{t}+c_2)}
\nonumber\\
& & \qquad\qquad\qquad\qquad\qquad\qquad \times \left[-d\hat{t}^2 + 
\sinh^2(7-p)\hat{t} dH_{8-p}^2\right]
\nonumber\\
& & + \left(\cosh\frac{\chi\alpha}{2}(\hat{t}-t_0)\right)^{-1/2}
\left(\frac{(7-p)b}{16\alpha}\right)^{-1/2} e^{\frac{7-p}{8}(c_1\hat{t}+c_2)}
dx_{(p+1)}^2
\eea
with the dilaton and the $(8-p)$-form retaining the same form as given in
(3.3). So unless we impose any further condition the SD$p$-brane solution
would depend on four parameters and will be different from the KMP solution.
We will now try to map this four parameter solution to the KMP solution
and see how the parameters in these two solutions are related.

In order to do this we make a coordinate transformation (1.1) i.e.
\be
\hat{t} = - \frac{1}{7-p} \ln \left(\frac{f_-}{f_+}\right)
\ee
where $f_{\pm}$ are defined as,
\be
f_{\pm} = 1 \pm \left(\frac{\omega}{t}\right)^{7-p}
\ee
Then we find,
\bea
& &\frac{1}{\left(\sinh(7-p)\hat{t}\right)^{\frac{2(8-p)}{(7-p)}}}
\left[-d\hat{t}^2 + \sinh^2 (7-p)\hat{t}\,\, dH_{8-p}^2\right]\nonumber\\
& & = \frac{\left(f_+ f_-\right)^{\frac{2}{(7-p)}}}{2^{\frac{2}{(7-p)}}\omega^2}
\left(-dt^2 + t^2 dH_{8-p}^2\right)
\eea
Also we rewrite
\bea
& & \frac{(7-p)b}{16\alpha}
e^{\frac{(p+1)}{4}(c_1\hat{t} + c_2)}
\cosh\frac{\chi\alpha}{2}(\hat{t}-t_0)\nonumber\\ 
& & = 2^{\frac{4}{(7-p)}} \omega^4 \left(\frac{f_-}{f_+}\right)^
{\frac{2n(p-1)}{(7-p)}}\Big[\cos^2\theta \left(\frac{f_-}{f_+}\right)^
{\frac{n-m}{2}} + \sin^2\theta \left(\frac{f_-}{f_+}\right)^{\frac{n+m}{2}}
\Big]\nonumber\\
& & = 2^{\frac{4}{(7-p)}} \omega^4 \left(\frac{f_-}{f_+}\right)^
{\frac{2n(p-1)}{(7-p)}}\,\, F 
\eea
where we have defined
\bea
\cos^2\theta &=& \frac{1}{2} e^{-\frac{\chi\alpha}{2}t_0}\left(\frac{(7-p)b}
{16\alpha}\right) e^{\frac{(p+1)(7-p)}{4(3-p)}c_2}\nonumber\\
\sin^2\theta &=& \frac{1}{2} e^{\frac{\chi\alpha}{2}t_0}\left(\frac{(7-p)b}
{16\alpha}\right) e^{\frac{(p+1)(7-p)}{4(3-p)}c_2}
\eea
We find that eq.(3.10) will be consistent if $m$, $n$ and $\omega$ satisfy
the following relations,
\be
m = \frac{32\alpha}{(7-p)^2},\qquad n = -\frac{c_1}{6},\qquad
\omega = \frac{1}{2^{\frac{1}{(7-p)}}} e^{\frac{(p+1)}{4(p-3)} c_2}
\ee
Now using (3.9) and (3.10) the metric (3.6) takes the form,
\bea
ds^2 &=& F^{1/2} \left(\frac{f_-}{f_+}\right)^{\frac{n(p-1)}{(7-p)}}
\left(f_+ f_-\right)^{\frac{2}{7-p}}\left(-dt^2 + t^2 dH_{8-p}^2\right)
\nonumber\\
& & \qquad\qquad\qquad\qquad + F^{-1/2} \left(\frac{f_-}{f_+}\right)^n
dx_{(p+1)}^2
\eea
This is precisely the form of the metric for the SD$p$-brane obtained in 
\cite{kmp}. In writing (3.13) we have rescaled the coordinates $x_i$,
for $i = 1, 2, \ldots, p+1$, by $x_i \to (2\omega^{7-p})^{1/(p+1)}x_i$. Also
from (3.11) we find 
\bea
\tan\theta &=& e^{\frac{16 \alpha t_0}{(7-p)}}\nonumber\\
c_2 &=& - \frac{4(3-p)}{(p+1)(7-p)}\ln \left(\frac{(7-p)b}{16\alpha}\cosh
\frac{16\alpha t_0}{(7-p)}\right)
\eea
We thus find that the parameter $c_2$ is determined in terms of $\alpha$,
$b$ and $t_0$ and so, we are left with four parameter solution $\alpha$,
$b$, $c_1$, $t_0$ with a relation between $c_1$ and $\alpha$ given in (3.5),
just like the KMP solution which is dependent on four parameters $m$, $n$,
$\omega$ and $\theta$ with a relation between $m$ and $n$ \cite{kmp}. 
It should be
emphasized that unlike the other conditions given in (3.12) and (3.14),
the second relation of (3.14) does not relate the CGG parameters with the
KMP parameters. Rather it is a relation among the parameters of the CGG 
solution itself and this reduces the number of parameters from four to three.
The reason behind this is while trying to map the CGG solution to the KMP
solution we are imposing the boundary condition (i.e. the metric becomes flat
and $e^{2\phi}$ approaches unity as $\hat{t} \to 0$) of the KMP solution here.
 Note that using
(3.12) and (3.5) we get the relation between $m$ and $n$ as,
\be
9(p+1) n^2 + (7-p) m^2 = 8(8-p)
\ee
which is the same as eq.(15) in ref.\cite{kmp}. The relations between the 
CGG parameters 
and the KMP parameters are
\bea
m &=& \frac{32\alpha}{(7-p)^2}\nonumber\\
n &=& - \frac{c_1}{6}\nonumber\\
\omega &=&\Big[\frac{(7-p)b}{32\alpha} \cosh \frac{16\alpha t_0}{(7-p)}\Big]^
{1/(7-p)}\nonumber\\
\tan\theta &=& e^{\frac{16\alpha t_0}{(7-p)}}
\eea
or, inverting the relations we get,
\bea
\alpha &=& \frac{(7-p)^2 m}{32}\nonumber\\
c_1 &=& - 6 n\nonumber\\
t_0 &=& \frac{2}{m(7-p)}\ln \tan\theta\nonumber\\
b &=& 2 (7-p) m \omega^{7-p} \sin\theta\cos\theta
\eea
Also using (3.10) and (3.11) we can write $e^{2\phi}$ in (3.3) precisely in
the same form as given in \cite{kmp}, i.e.,
\be
e^{2\phi} = F^{(3-p)/2}\left(\frac{f_-}{f_+}\right)^{pn}
\ee
and finally by taking a Hodge duality on $F_{8-p}$ given in (3.3) we find
\bea
\ast F_{p+2} &=& -e^{(p-3)\phi/2} e^{2f(p+1)} b \, d\hat{t}\wedge dx_1\wedge
\ldots\wedge dx_{p+1}\nonumber\\
&=& 2(7-p) m \sin\theta\cos\theta \omega^{7-p} t^{p-8} \frac{1}{F^2}
\frac{1}{f_+ f_-}\left(\frac{f_-}{f_+}\right)^n dt\wedge dx_1\wedge\ldots
\wedge dx_{p+1}\nonumber\\
&=& \sin\theta\cos\theta \frac{d}{dt}\left(\frac{C}{F}\right) dt\wedge dx_1
\wedge\ldots\wedge dx_{p+1}
\eea
where $C = (f_-/f_+)^{(n+m)/2} - (f_-/f_+)^{(n-m)/2}$.
This is the form of field strength given in \cite{kmp}. In obtaining
(3.19) we have used the scaling of KMP coordinates $x_i$, $i=1,\ldots,p+1$
as $x_i \to (2\omega^{7-p})^{1/(p+1)} x_i$.

We now show the mapping of the CGG solution and the KMP solution for 
SD3-brane. As we have already mentioned that $\beta$ can be eliminated in
this case by just renaming the parameters $\alpha$, $t_0$, $c_1$ as in
(3.4) along with the coordinate rescaling $\hat{t} \to \hat{t}/\beta$ and
$x_i \to x_i/\beta^{1/4}$ for $i=1,2,3,4$. Note
from (3.3) that if we demand $e^{2\phi} \to 1$ as $\hat{t} \to 0$ in this case
then $c_2$ must vanish. The SD3-brane solution of CGG then reduces to,
\bea
ds^2 &=& \left(\frac{1}{\sinh 4\hat{t}}\right)^{5/2} \left(\cosh\frac{\chi
\alpha}{2}(\hat{t}-t_0)\right)^{1/2} \left(\frac{b}{4\alpha}\right)^{1/2}
e^{c_1\hat{t}/2}
\left[-d\hat{t}^2 + \sinh^2 4\hat{t} dH_5^2\right]\nonumber\\
& & \qquad\qquad\qquad\qquad + \left(\cosh\frac{\chi\alpha}{2}(\hat{t}-t_0)
\right)^{-1/2} \left(\frac{b}{4\alpha}\right)^{-1/2} e^{c_1\hat{t}/2} 
dx_{(4)}^2\nonumber\\
e^{2\phi} &=& e^{2c_1\hat{t}}\nonumber\\
F_5 &=& \frac{b}{\sqrt{2}}\left(1+\ast\right)\epsilon(H_5)
\eea
In the above $\chi = 8$ and the parameters $c_1$ and $\alpha$ satisfy
\be
c_1^2 + 16\alpha^2 = 40
\ee
Also the five-form field strength is self-dual.

Now with the coordinate transformation (3.7) and (3.8) with $p=3$ we get
\be
\left(\frac{1}{\sinh 4\hat{t}}\right)^{5/2}
\left[-d\hat{t}^2 + \sinh^2 4\hat{t}\,\, dH_5^2\right]
 = \frac{\left(f_+ f_-\right)^{1/2}}{2^{1/2}\omega^2}
\left(-dt^2 + t^2 dH_5^2\right)
\ee
and we also rewrite
\bea
\cosh\frac{\chi\alpha}{2}(\hat{t}-t_0) \left(\frac{b}{4\alpha}\right)
e^{c_1\hat{t}}
&=& 2 \omega^4 \left(\frac{f_-}{f_+}\right)^
n \Big[\cos^2\theta \left(\frac{f_-}{f_+}\right)^
{\frac{n-m}{2}} + \sin^2\theta \left(\frac{f_-}{f_+}\right)^{\frac{n+m}{2}}
\Big]\nonumber\\
&=& 2 \omega^4 \left(\frac{f_-}{f_+}\right)^
n\,\, F 
\eea
where we have defined
\bea
2\omega^4 \cos^2\theta &=& \frac{1}{2} e^{-\frac{\chi\alpha}{2}t_0}
\left(\frac{b}
{4\alpha}\right) \nonumber\\
2\omega^4 \sin^2\theta &=& \frac{1}{2} e^{\frac{\chi\alpha}{2}t_0}
\left(\frac{b}
{4\alpha}\right) 
\eea
Again from the consistency of eq.(3.23) and also from the relation
(3.24) we obtain the relations between the parameters $m$, $n$, $\omega$,
$\theta$ of KMP solution and $\alpha$, $c_1$, $t_0$, $b$ of CGG solution in
the same form as obtained in eq.(3.16) and eq.(3.17) with $p=3$. With these
we find that the metric and dilaton given in eq.(3.20) reduce to,
\bea
ds^2 &=& F^{-1/2} \left(\frac{f_-}{f_+}\right)^{n/2} 
\left(f_+ f_-\right)^{1/2}\left(-dt^2 + t^2 dH_5^2\right) + F^{-1/2} \left(
\frac{f_-}{f_+}\right)^n dx_{(4)}^2\nonumber\\
e^{2\phi} &=& \left(\frac{f_-}{f_+}\right)^{3n}
\eea
which have the same form as in \cite{kmp}. In the above we have 
rescaled the coordinates $x_{1,2,3,4} \to (2\omega^4)^{1/4} x_{1,2,3,4}$.
Finally, $F_5$ can be written in KMP parameters as,
\be
F_5 = \frac{1}{\sqrt{2}} 8 m \omega^4 \sin\theta\cos\theta (1+ \ast) 
\epsilon(H_5)
\ee
Thus we have shown the complete mapping of SD$p$-brane solutions of type
II string theory obtained by CGG and
KMP.

\sect{Conclusion}

To summarize, we have shown in this paper the complete mapping of SD$p$-brane 
solutions of type II string theory obtained by Chen-Gal'tsov-Gutperle and
Kruczenski-Myers-Peet. After carefully eliminating some of the parameters
we have indicated that the solution of CGG is a four parameter solution and
therefore is more general than the three parameter solution of KMP. In 
contrast to the CGG solution, KMP solution uses a specific boundary condition
of the metric and the dilaton. We have shown that when the same boundary
condition of the KMP solution is imposed upon the CGG solution, then one of
the parameters of the CGG solution is removed and therefore both the solutions
become three parameter solutions. Only under this condition both the solutions
become identical and are related by a coordinate transformation as well
as a Hodge duality of the field strength. The three parameters of both the
solutions are related non-trivially to one another and we have explicitly given
these relations. Since space-like D$p$-branes, particularly the supergravity
descriptions are important to understand the (time-like) holography of dS/CFT
correspondence, we hope that the equivalence of two supergravity descriptions
of SD$p$-branes shown in this paper will be helpful for this purpose as well 
as to understand the physical properties of these unusual branes.

\section*{Acknowledgements}

I would like to thank Somdatta Bhattacharya for checking some of the
results in the paper. I would also like to thank L. Cornalba for pointing
out ref.\cite{cck} to me.

\end{document}